\documentclass{aa}
\usepackage{amssymb}
\usepackage[pdftex]{graphicx}
\usepackage{natbib}
\bibpunct{(}{)}{;}{a}{}{,}

\begin{document}
   \title{Coronal ion-cyclotron beam instabilities within the multi-fluid description}


   \author{R. Mecheri
          \inst{1}
          \and
           E. Marsch\inst{1}  
           }

   \offprints{R. Mecheri}

   \institute{Max-Planck Institut f\"{u}r Sonnensystemforschung,
              Max-Planck Strasse 2, 37191 Katlenburg-Lindau, Germany\\
              \email{mecheri@mps.mpg.de}
             }

    \date{Received xxxx; accepted xxxx}


  \abstract
   {Spectroscopic observations and theoretical models suggest
   resonant wave-particle interactions, involving high-frequency
   ion-cyclotron waves, as the principal mechanism for heating and accelerating
   ions in the open coronal holes. However,
   the mechanism responsible for the generation of the ion-cyclotron waves remains
   unclear. One possible scenario is that ion beams originating from small-scale
   reconnection events can drive micro-instabilities that constitute a possible
   source for the excitation of ion-cyclotron waves.}
   {In order to study ion beam-driven electromagnetic instabilities, the multi-fluid model
   in the low-$\beta$ coronal plasma is used. While neglecting
   the electron inertia this model allows one to take into account ion-cyclotron wave
   effects that are absent from the one-fluid MHD model. Realistic models of density
   and temperature as well as a 2-D analytical magnetic field model are used to define the
   background plasma in the open-field funnel region of a polar coronal hole.}
   {Considering the WKB approximation, a Fourier plane-wave linear mode analysis is
    employed in order to derive the dispersion relation. Ray-tracing theory is used to
    compute the ray path of the unstable wave as well as the evolution of the growth
    rate of the wave while propagating in the coronal funnel.}
   {We demonstrate that, in typical coronal holes conditions and assuming realistic
    values of the beam velocity, the free energy provided by the ion beam propagating
    parallel the ambient field can drive micro-instabilities through resonant
    ion-cyclotron excitation.}
   {}

\titlerunning{Coronal ion-cyclotron beam instabilities}
\authorrunning{R. Mecheri \& E. Marsch}
   \maketitle
%

\section{Introduction}\label{introduction}


Observations made by the Ultraviolet Coronagraph Spectrometer (UVCS)
and other instruments on the Solar and Heliospheric Observatory
(SOHO) have significantly increased our knowledge of the kinetic
properties of charged particles close to the Sun in the source
region of the fast solar wind. Spectroscopic determination of the
widths of ultraviolet emission lines in coronal holes indicate that
heavy ions are very hot and have high temperature anisotropies, and
that heavier ions have a higher temperature than the protons by at
least their mass ratio, i.e. $T_{i}/T_{p}>m_{i}/m_{p}$
\citep{Mecheri:Kohl,Mecheri:Cranmer}. These observations strongly
suggest resonant ion-cyclotron wave-particle interaction as a major
mechanism for the heating and acceleration of ions in the
magnetically open corona. This notion led to a renewed interest in
models involving ion heating by high-frequency ion-cyclotron waves
\citep{Mecheri:Isenberg00,Mecheri:Hollweg00,Mecheri:Marsch01,
Mecheri:Vocks01,Mecheri:Xie04}. For a detailed review on resonant
ion-cyclotron interactions in the corona, see
\citet{Mecheri:Hollweg02}.

%

However, it remains unclear how these waves originate in the solar
corona, and whether they are generated locally or emanate from the
coronal base. One possible scenario is that the ion-cyclotron waves
are generated in the lower corona by small-scale reconnection events
as suggested by \citet{Mecheri:Axford92}. They might also be
generated locally through a turbulent cascade of low-frequency
MHD-type waves towards high-frequency ion-cyclotron waves
\citep{Mecheri:Li99,Mecheri:Hollweg00,Mecheri:Ofman02}, or by plasma
microinstabilities driven by current fluctuations of low-frequency
MHD modes \citep{Mecheri:Markovskii01}, or by an intermittent
electron heat flux accompanying microflare events
\citep{Mecheri:Markovskii04}. \citet{Mecheri:Voitenko02} further
suggested that plasma outflows from reconnection sites in
microflares could create ion-beam configurations in the surrounding
plasma, and thus provide free energy for driving kinetic
microinstabilities through ion-cyclotron resonance and the Cerenkov
effect. The possible origin of the ion beams observed in the solar
wind from reconnection jets and explosive events in the corona has
been proposed by \citet{Mecheri:Feldman}.

Indeed, detailed spectroscopic studies of the so-called
high-velocity events and explosive events, using spectra obtained
with the Coronal Diagnostic Spectrometer (CDS;
\cite{Mecheri:Brekke}) and the Solar Ultraviolet Measurement of
Emitted Radiation instrument (SUMER; \cite{Mecheri:Innes}) both on
SOHO, revealed a new character of the lower corona as a highly
dynamic medium. They signify the omni-presence of transient
explosive events and a wide variety of plasma jets with velocities
ranging from a few tens of a kilometer per second up to several
hundreds of kilometers per second. Since these plasma jets have been
observed to evolve in a similar way as predicted by the theory of
magnetic reconnection \citep{Mecheri:Innes}, explosive events and
plasma jets have been associated with the highly-dynamic small-scale
reconnections which are supposed to take place in the chromospheric
network, approximately at heights of $1000-3000$~km above the
photosphere. Solar magnetograms provided by the Michelson Doppler
Imager (MDI) on SOHO clearly indicate that the magnetic network
field exists in two characteristic components, i.e. in few unipolar
open flux tubes, "funnels", and in multiple closed flux tubes,
"loops", \citep{Mecheri:Schrijver98}. The magnetic network is very
dynamic and releases non-potential-field magnetic energy which is
converted into plasma heating, beam particles, and motion of the
coronal plasma. All these processes provide ample free energy for
driving plasma macro- and micro-instabilities.

The present paper aims at studying the possible occurrence of these
instabilities in typical conditions for coronal holes. We
particularly focus our study on beam-driven instabilities in the
ion-cyclotron frequency range which are driven by the presence of
tenuous ion beams, presumably originating from small-scale
reconnection sites and propagating parallel to the ambient magnetic
field. Linear mode analysis is used in the framework of the
multi-fluid model, while neglecting the electron inertia. This model
permits the consideration of ion-cyclotron-wave effects that are
absent from the one-fluid MHD model. Realistic models of the density
and temperature, as well as a 2-D funnel model describing the
open-field region, are used to define the background plasma.
Considering the WKB approximation (in which the wavelength of
interest is assumed to be smaller than the non-uniformity length
scale), we first solve the dispersion relation locally and then
perform a non-local wave analysis using the ray-tracing theory,
which allows us to compute the ray paths of the unstable waves in
the funnel as well as the spatial variation of their growth rates.

This paper is structured as follows. In Sect.~\ref{background}, we
present the 2-D analytical funnel model used in this study to
describe open-field region in a coronal hole. Then in
Sect.~\ref{linear}, we describe how the local and non-local (ray
tracing) linear perturbation analysis using the multi-fluid model is
carried out. The results are presented and discussed in
Sect.~\ref{results}, and finally we give our conclusions in
Sect.~\ref{conclusion}.

\section{Background plasma configuration}\label{background}

\begin{figure}
\centering $\begin{array} {c@{\hspace{0.05in}}c}
\includegraphics[width=6.8cm]{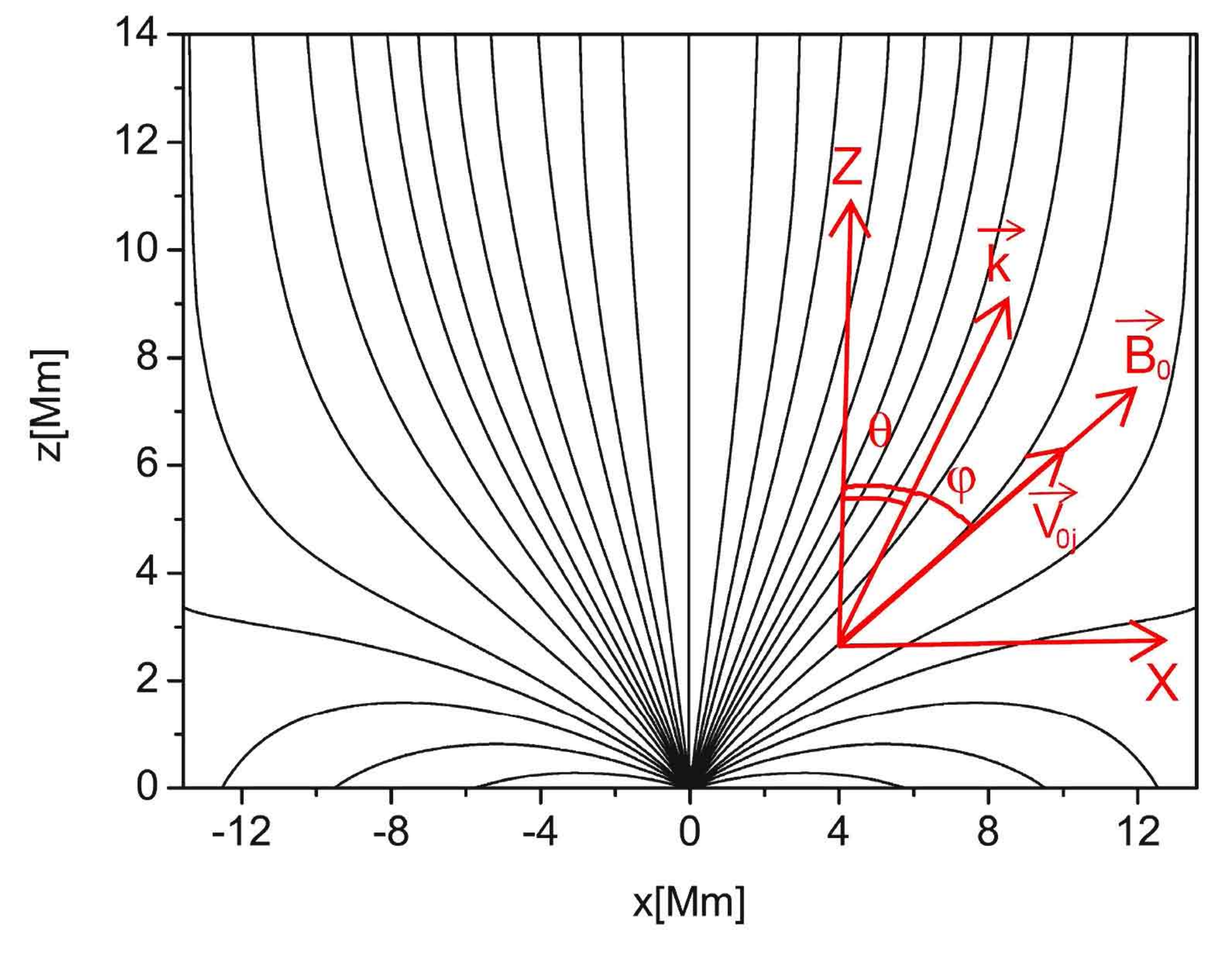}
\end{array}$
\vspace{-0.25cm}\caption{
Magnetic field geometry of a funnel as obtained from the 2-D
potential-field model derived by \citet{Mecheri:HackenbergB}. The
field lines emerge from the boundary between two adjacent
supergranules ($x=0$) and expand rapidly to fill the corona. The
photospheric level is at $z=0$. Coordinate axes, wave vector and
beam drift velocity are shown in red color.} \label{Figfunnel}
\end{figure}

For the background plasma density and temperature we use the model
parameters of \citet{Mecheri:Fontenla} for the chromosphere and
\citet{Mecheri:Gabriel} for the lower corona. The 2-D
potential-field model derived by \citet{Mecheri:HackenbergB} is used
here to define the background magnetic field (Fig. \ref{Figfunnel})
in the funnel. Analytically, the two components of this model field
are given by: {\setlength\arraycolsep{0.005em}
\begin{eqnarray}
B_{0x}(x,z)=\frac{(B_{max}-B_{00})L}{2\pi(L-d)}ln\frac{cosh\frac{2\pi
z}{L}- cos(\frac{\pi d}{L}+\frac{2\pi x}{L})}{cosh\frac{2\pi z}{L}-
cos(\frac{\pi d}{L}-\frac{2\pi x}{L})}\nonumber\\
\end{eqnarray}
\vspace{-0.7cm}
\begin{eqnarray}
B_{0z}(x,z)&=&B_{00}+(B_{max}-B_{00})\left[-\frac{d}{L-d}+\frac{L}{(L-d)\pi}\right.
\times\nonumber\\
&&\left(arctan\frac{cosh\frac{2\pi z}{L}~sin\frac{\pi d}{2L}+
sin(\frac{\pi d}{2L}+\frac{2\pi x}{L})}{sinh\frac{2\pi
z}{L}~cos\frac{\pi d}{2L}}+\right.
\nonumber\\
&&\left.\left.arctan\frac{cosh\frac{2\pi z}{L}~sin\frac{\pi d}{2L}+
sin(\frac{\pi d}{2L}-\frac{2\pi x}{L})}{sinh\frac{2\pi
z}{L}~cos\frac{\pi d}{2L}}\right)\right]
\end{eqnarray}}
The typical parameters relevant for this model are: $L=30~
\textrm{Mm}, d=0.34~\textrm{Mm}, B_{00}=11.8~\textrm{G}, $ and
$B_{max}=1.5~\textrm{kG}$.

\section{Linear perturbation analysis}\label{linear}

To describe wave propagation in the funnel we use the multi-fluid
equations and subject them to a linear perturbation analysis. The
fluid equations associated with the polytropic gas law for any
particle species $j$ are given by:
\begin{eqnarray}
  \frac{\partial n_{j}}{\partial
  t}+\nabla\cdot(n_{j}\textbf{v}_{j})=0,
  \label{mass}
\end{eqnarray}
\begin{eqnarray}
  m_{j}n_{j}(\frac{\partial \textbf{v}_{j}}{\partial
  t}&+&\textbf{v}_{j}\cdot\nabla \textbf{v}_{j})+
  \nabla p_{j}-q_{j}n_{j}(\textbf{E}+\textbf{v}_{j}\times
  \textbf{B})+\nonumber\\&+&
  m_{j}n_{j}\sum_{j'}\nu_{jj'}(\textbf{v}_{j}-\textbf{v}_{j'})=
  \mathbf{0},
  \label{momentum}
\end{eqnarray}
\begin{equation}
  p_{j}n_{j}^{-\gamma_{j}}=\textrm{const},
  \label{energy}
\end{equation}
where $m_{j}$, $n_{j}$, $\textbf{v}_{j}$, $p_{j}$ and
$\gamma_{j}$(=5/3) are respectively the mass, density, velocity,
pressure and the adiabatic polytropic index of a species $j$.
Subscript $j$ stands for electron $e$, proton $p$ or alpha particle
$\alpha$ (He$^{2+}$). The quantity $\nu_{jj'}$ is the collision
frequency of a particle of species $j$ with particles of species
$j'$ (only the electron-proton collisions are taken into account).
The electric field $\textbf{E}$ and the magnetic field $\textbf{B}$
are linked by Faraday's law:
\begin{equation}
\nabla\times \textbf{E}=-\frac{\partial \textbf{B}}{\partial t}.
\end{equation}

\subsection{Linearization procedure}

The linear perturbation analysis is performed by expressing all the
quantities in the fluid equations as a sum of an unperturbed
stationary part (with subscript 0) and a perturbed part (with
subscript 1) that is much smaller than the stationary part:
{\setlength\arraycolsep{-0.05cm}
\begin{eqnarray}
&n_{j}&=n_{0j}(z)+n_{1j},~T_{j}=T_{0j}(z)+T_{1j},~p_{j}=p_{0j}(z)+p_{1j},
\nonumber\\[0.25cm]
&\textbf{v}_{j}&=\textbf{v}_{0j}+\textbf{v}_{1j},~
\textbf{B}=\textbf{B}_{0}(x,z) +\textbf{B}_{1},~
\textbf{E}=\textbf{E}_{0}+\textbf{E}_{1}, \nonumber\\\nonumber\\[-0.05cm]
&\textrm{with:}&~~n_{1j}\ll n_{0j}, ~~T_{1j}\ll T_{0j},~~p_{1j}\ll
p_{0j}, ~~\left| \textbf{v}_{1j} \right| \ll \left| \textbf{v}_{0j}
\right|,
\nonumber\\[0.25cm]
&\left|\textbf{B}_{1}\right|&\ll\left|\textbf{B}_{0}\right|,
~~\left|\textbf{E}_{1}\right|\ll\left|\textbf{E}_{0}\right|
~~~~\textrm{and}~~~~\textbf{v}_{0j}\times\textbf{B}_{0}=\textbf{0}.
\label{perturbation}
\end{eqnarray}}\\
We assume charge neutrality and a current-free state for the
unperturbed stationary plasma, i.e., $\sum_{j}q_{j}n_{0j}=0$ and
$\sum_{j}q_{j}n_{0j}\textbf{v}_{0j}=0$. The zero-order terms cancel
out when Eq. (\ref{perturbation}) is inserted into the multi-fluid
Eqs. (\ref{mass})-(\ref{energy}). Neglecting the nonlinear products
of the first-order terms, we get a system of coupled linear
equations:

\begin{equation}
i(\omega-\textbf{k}\cdot\textbf{v}_{0j})\frac{n_{1j}}{n_{0j}}
-i\textbf{k}\cdot\textbf{v}_{1j}=0,
\end{equation}
{\setlength\arraycolsep{-0.06cm}
\begin{eqnarray}
&i&(\omega-\textbf{k}\cdot\textbf{v}_{0j})\textbf{v}_{1j}+
\Omega_{j}\left(\frac{\textbf{E}_{1}}{\left|\textbf{B}_{0}\right|}
+\textbf{v}_{1j}\times\frac{\textbf{B}_{0}}{\left|\textbf{B}_{0}\right|}+
\textbf{v}_{0j}
\times\frac{\textbf{B}_{1}}{\left|\textbf{B}_{0}\right|}
\right)\nonumber\\
&-& i C_{sj}^{2}\frac{p_{1j}}{p_{0j}}\textbf{k}-
\sum_{j'}\nu_{jj'}(\textbf{v}_{1j}-\textbf{v}_{1j'})
=\mathbf{0},
\end{eqnarray}}
\begin{equation}
\frac{p_{1j}}{p_{0j}}-\gamma_{j}\frac{n_{1j}}{n_{0j}}=0,
\end{equation}
\begin{equation}
i\textbf{k}\times \textbf{E}_{1}=i\omega \textbf{B}_{1},
\end{equation}
where all the perturbed quantities have been expressed in form of a
plane wave. This Fourier analysis turns all derivatives into
algebraic factors, i.e. $\partial/\partial t\rightarrow -i \omega$
and $\nabla\rightarrow i \textbf{k}$, where $\omega$ is the wave
frequency and $\textbf{k}$ the wave vector. In the above equations,
$C_{sj}=\sqrt{\gamma_{j}k_{B}T_{j}/m_{j}}$ is the acoustic speed and
$\Omega_{j}=eB_{0}/m_{j}$ the cyclotron frequency of species $j$,
and $k_{B}$ is the Boltzman constant.

\subsection{Dispersion relation}

To derive the dispersion relation, the above linearized equations
have to be combined in order to obtain a linear relation between the
current density $\textbf{J}_{1}$ and the electric field
$\textbf{E}_{1}$:

\begin{equation}
\textbf{J}_{1}=\vec{\sigma}\cdot\textbf{E}_{1},
\end{equation}
where $\vec\sigma$ is the conductivity tensor which is related to
the dielectric tensor $\vec\epsilon$ through the following relation:

\begin{equation}
\vec\epsilon(\omega,\textbf{k},\textbf{r})=\textbf{I}+\frac{i}
{\omega\varepsilon_{0}}\vec\sigma(\omega,\textbf{k},\textbf{r}).
\label{dielectric}
\end{equation}
Finally, the local dispersion relation is obtained using the theory
of electrodynamics \citep[see, e.g.,][]{Mecheri:Stix}:
\begin{equation}
 D(\omega,\textbf{k},\textbf{r})=\textrm{Det}\left[\frac{c^{2}}{\omega^{2}}
 \textbf{k}\times(\textbf{k}\times\textbf{E})+\vec{\epsilon}(\omega,
 \textbf{k},\textbf{r})\cdot\textbf{E}\right]=0,
\label{dispersion}
\end{equation}
where $c$ is the speed of light in vacuum and $\textbf{r}$ is the
large-scale position vector. We choose the wave vector $\textbf{k}$
to lie in the $x-z$ plane, with
$\textbf{k}=k(sin\theta,0,cos\theta)$.

\begin{figure*}
\centering $\begin{array} {c@{\hspace{0.05in}}c} \multicolumn{1}{c}
{\hspace{0.5cm}\mbox{\Large$x=7.5$~Mm,~$z=2.2$~Mm,~$\varphi\approx82^{\circ}$}}
\\[-0.cm]
\includegraphics[width=14.5cm]{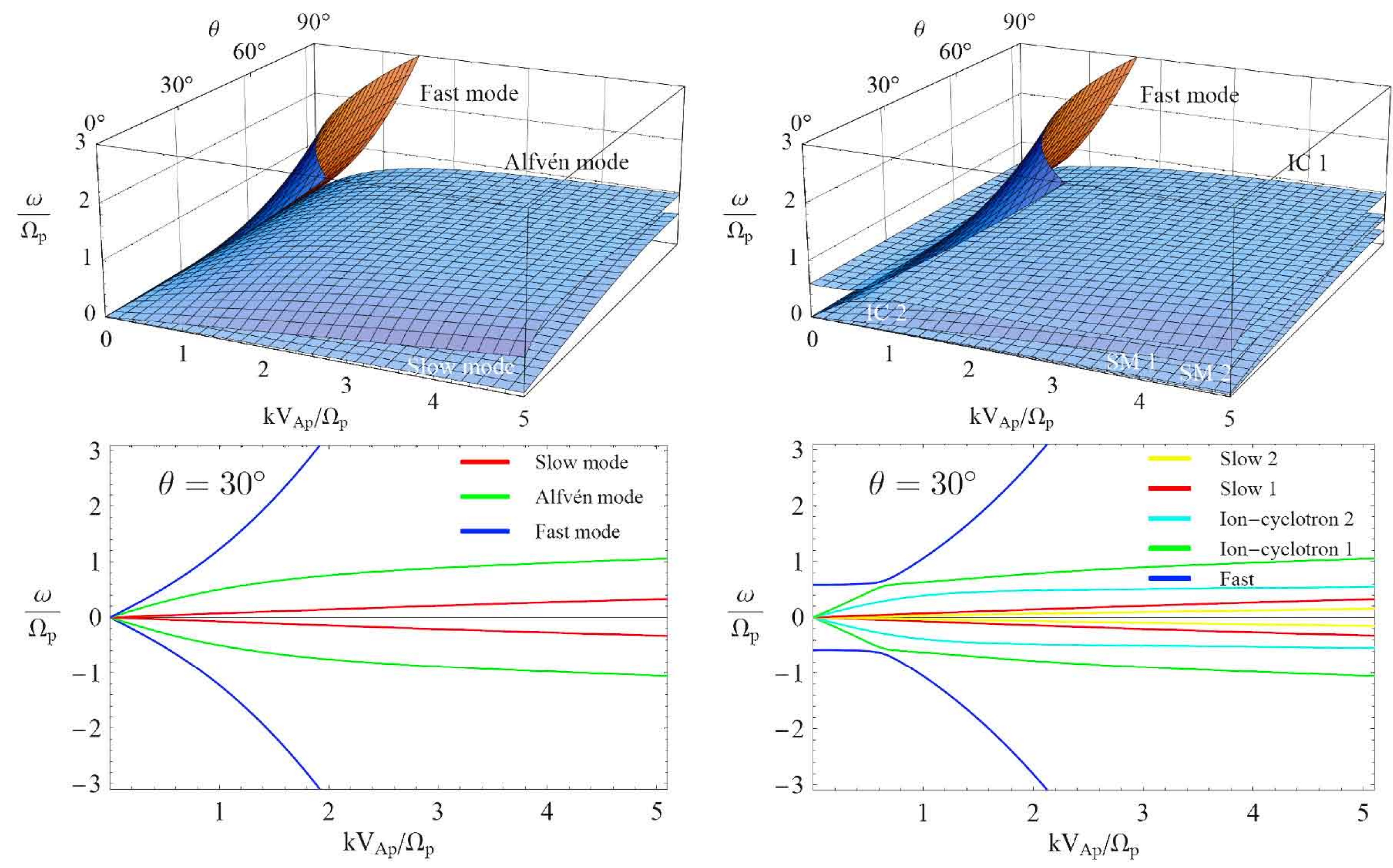}
\end{array}$
\vspace{-0.cm} \caption{Two-fluid (e-p plasma on the left panels)
and three-fluid (e-p-He$^{2+}$ plasma on the right panels)
dispersion surfaces (top panels) and single curves (for
$\theta=30^{\circ}$ at the bottom) for the case of a zero beam
speed, i.e. $v_{0\alpha}=0$. The alpha-particle (He$^{2+}$) density
is $n_{0\alpha}=0.1n_{0p}$ and the plasma beta values are
$\beta_{e}=0.0043$ = 1.2$\beta_{p}=12\beta_{\alpha}$. Here $\omega$
and $k$ are normalized, respectively, to the proton cyclotron
frequency, $\Omega_{p}$, and the inertial length,
$\Omega_{p}/V_{Ap}$, where
$V_{Ap}$=$\textrm{B}_{0}$/$\sqrt{\mu_{0}n_{0p}m_{p}}$ is the proton
Alfv\'{e}n speed ($\mu_{0}$ is the magnetic permeability in the
vacuum). Here $T_{0e}=T_{0p}=T_{0\alpha}=1.35\times10^{4} K$,
$n_{0p}=3.73\times10^{16}m^{-3}$, $\textrm{B}_{0}=15.54$ Gauss.}
\label{23Dr23}
\end{figure*}

\subsection{Ray-tracing equations}

In the framework of the WKB approximation, the ray tracing problem
consists in solving a system of ordinary differential equations of
the Hamiltonian form \citep{Mecheri:Weinberg}. The ray-tracing
equations, which represent the equations of motion for the wave
frequency $\omega$, the wave vector $\textbf{k}$, and the space
coordinate $\textbf{r}$, have been formulated by
\citet{Mecheri:Bernstein}. In the simple case of a hermitian
dielectric tensor, they are given by:

\begin{equation}
\frac{\textrm{d}\omega}{\textrm{d}t}=-\frac{\partial
D(\omega,\textbf{k} ,\textbf{r})/\partial t}{\partial
D(\omega,\textbf{k} ,\textbf{r})/\partial \omega}=0, \label{rt1}
\end{equation}
\begin{equation}
\frac{\textrm{d}\textbf{k}}{\textrm{d}t}=~\frac{\partial
D(\omega,\textbf{k} ,\textbf{r})/\partial \textbf{r}}{\partial
D(\omega,\textbf{k} ,\textbf{r})/\partial \omega}, \label{rt2}
\end{equation}
\begin{equation}
\frac{\textrm{d}\textbf{r}}{\textrm{d}t}=-\frac{\partial
D(\omega,\textbf{k} ,\textbf{r})/\partial \textbf{k}}{\partial
D(\omega,\textbf{k} ,\textbf{r})/\partial \omega}. \label{rt3}
\end{equation}
A generalization of these equations to the case of an anti-hermitian
dielectric tensor was also proposed in the paper of
\citet{Mecheri:Bernstein}. In this case, in addition to the ray
path, the growth rate of the instability can be computed as well.
Note that Eq.~(\ref{rt1}) can be set to zero, because the dispersion
relation does not explicitly depend on the time $t$ (the background
plasma is stationary). The above set of differential equations
represents an initial-value problem which can be solved by using the
initial conditions obtained from the local solutions of the
dispersion relation (Eq.~\ref{dispersion}).
\section{Numerical Results}\label{results}
We assume that the ion-beam particles are generated at the funnel
location at x = 7.5 Mm and z = 2.2 Mm, presumably by small scales
reconnection events. This location is characterized by a magnetic
field inclination angle of $\varphi=82^{\circ}$ with respect to the
normal on the solar surface. According to the observations, the
calculations will be performed for a beam velocity equal to
320~km/s. We consider the case of an alpha-particle (He$^{2+}$) beam
plasma configuration, propagating parallel to the ambient field. For
comparison purpose, the dispersion diagrams in the case of a plasma
without the beam are also presented. We first present the results
obtained from the local solutions of the dispersion relation
(\ref{dispersion}) and then the results obtained from the non-local
wave analysis using the ray-tracing equations.

\subsection{Local stability analysis}

\subsubsection{Zero beam plasma}

For $v_{0\alpha}=0$ and $\theta=30^{\circ}$, the dispersion diagram
in the case of the two-fluid (e-p) model (left panels of Fig.
\ref{23Dr23}) shows the presence of three stable modes, each one of
them is represented by an oppositely propagating ($\omega>0$ and
$\omega<0$) pair of waves. These modes represent the extensions of
the usual Slow (red line), Alfv\'{e}n (green line) and Fast (blue
line) MHD modes into the high-frequency domain around
$\omega=\Omega_{p}$ ($=eB_{0}/m_{p}$), where the waves are
dispersive \citep[for details see][]{Mecheri:Mecheri}. In the case
of the three fluid (e-p-He$^{2+}$) model (right panels of
Fig.~\ref{23Dr23}) five stable modes are present (each one
represented by an oppositely propagating pair of waves). These modes
are: two slow modes (yellow and red lines), two ion-cyclotron modes
(gray and green lines) and one fast mode (blue line). The waves are
subject to mode conversion or coupling, a phenomenon associated with
the appearance of a cut-off frequency concerning the fast mode
\citep[for details see][]{Mecheri:Mecheri}.

\subsubsection{Alpha particles beam}

The alpha particle beam configuration consists of three species:
electrons (with density $n_{0e}$), protons (with density $n_{0p}$)
and a tenuous beam of alpha particles, He$^{2+}$ indicated by
$\alpha$, with a velocity $v_{0\alpha}$ parallel to the ambient
magnetic field $\textbf{B}_{0}$ and a density
$n_{0\alpha}=0.1n_{0p}$. The protons are considered to be at rest
and the electrons are in motion with a velocity $v_{0e}$. Alphas and
electrons satisfy thus the zero-current condition,
$v_{0e}=2(n_{0\alpha}/n_{0e})v_{0\alpha}$. The alpha-particle-beam
relative density is taken from in-situ observations made by the
Helios spacecraft, with $n_{0\alpha}/n_{0e}\approx
n_{0\alpha}/n_{0p}=0.05-0.2$ \citep{Mecheri:Tu04}. This choice is
justified since it has been argued by \citet{Mecheri:Feldman} that
the alpha beams might originate from reconnection events at the base
of the expanding solar corona.
\begin{figure}
\centering $\begin{array}{c@{\hspace{0.05in}}c}
\multicolumn{1}{c}{\mbox{\large
x=7.5~Mm,~z=2.2~Mm,~$\varphi\approx82^{\circ}$}}
\end{array}$
\\[-0.cm]
$\begin{array} {c@{\hspace{0.05in}}c}
\includegraphics[width=8.7cm]{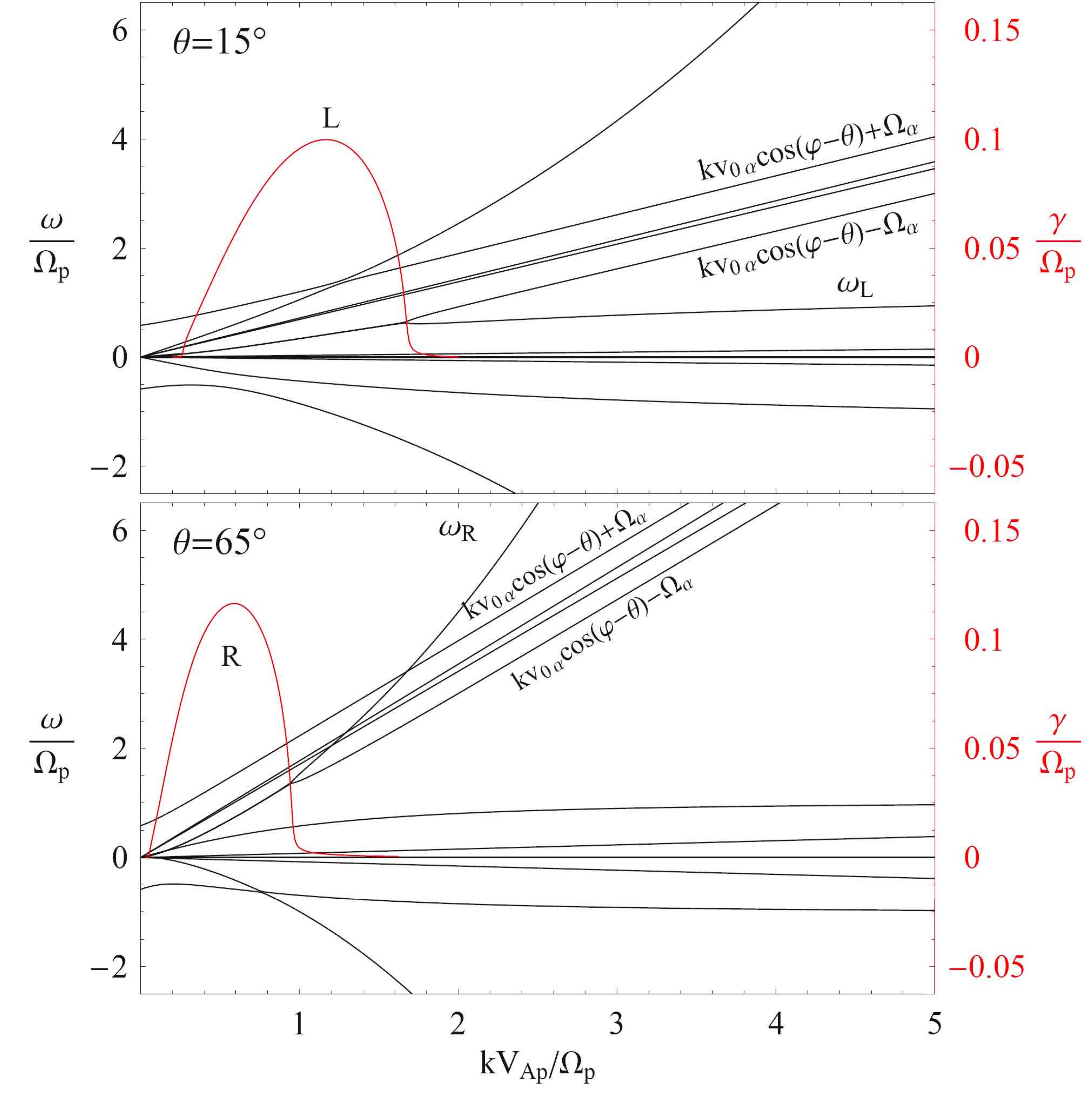}
\end{array}$
\vspace{-0.4cm}\caption{Wave frequency (black) and growth rate (red)
in a alpha particle beam plasma ($n_{0\alpha}/n_{0p}=0.1$ and
$v_{0\alpha}=320~\textrm{km/s} \approx 1.8 V_{Ap}$) versus wave
vector for two propagation angles $\theta$. Note, that for
$\theta=15^{\circ}$ (i.e., quasi-perpendicular propagation, top
panel), the left-hand resonant ion-cyclotron mode ($\omega_{L}$) is
excited, and that for $\theta=65^{\circ}$ (i.e., quasi-parallel
propagation, bottom panel) the right-hand resonant fast mode
($\omega_{R}$) is excited, whereby both satisfy the condition
(\ref{ano-dop}) for the anomalous Doppler effect. Here
$T_{0e}=T_{0p}=T_{0\alpha}$.} \label{2D-LR-al}
\end{figure}
In this case (i.e. $v_{0b}\neq$0) Fig.~\ref{2D-LR-al} shows that the
dispersion curves are strongly modified by the appearance of regions
of instability associated with two cyclotron beam modes, which in
the cold-plasma case are characterized by the dispersion relations
\citep{Mecheri:Cap}:

\begin{equation}
\omega_{b} \approx \left\{ \begin{array}{ll}
kv_{0\alpha}~cos(\varphi-\theta)+\Omega_{\alpha} &
~~\textrm{cyclotron
beam 1}\\
kv_{0\alpha}~cos(\varphi-\theta)-\Omega_{\alpha} &
~~\textrm{cyclotron beam 2}
\end{array} \right.
\label{beam-modes}
\end{equation}

Therefore, the value of electron-proton collision frequencies in
this region of the solar atmosphere (i.e.
$\nu_{ep}/\Omega_{p}\approx$11.6 and
$\nu_{pe}/\Omega_{p}\approx$0.006) are not large enough to prevent
from the instabilities to develop. What collisions do is to make the
electron and proton bulk velocity equal, but the alpha differential
motion (which provide the free energy) remain unaffected because of
the high beam velocity considered.

\begin{figure*}
\begin{center}
$\begin{array} {c@{\hspace{0.05in}}c}
\multicolumn{1}{c}{\mbox{\Large $x=7.5$~Mm,~$z=2.2$~Mm,
~$\varphi\approx82^{\circ}$}}\\[0.1cm]
\includegraphics[width=16.cm]{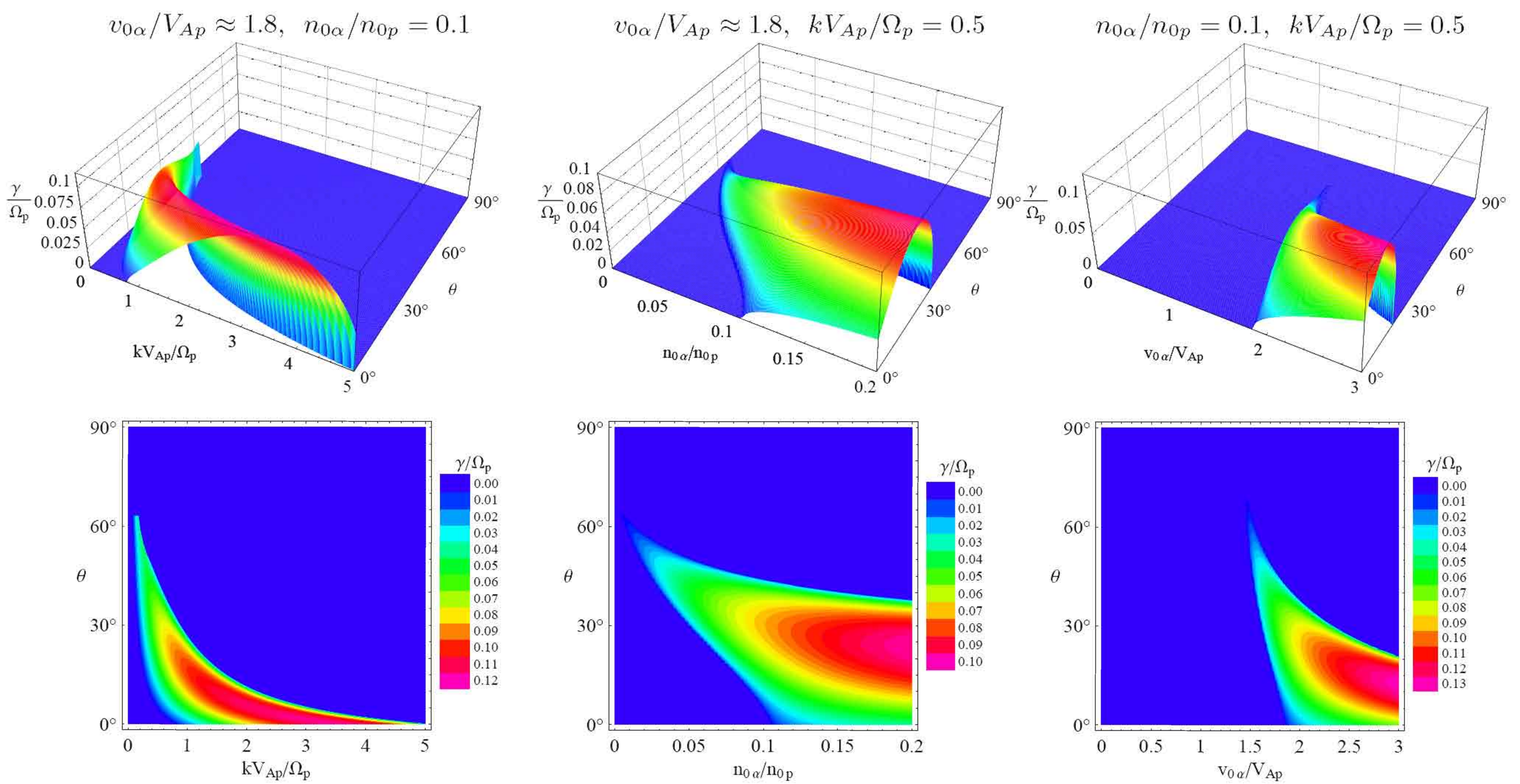}
\end{array}$
\end{center}
\vspace{-0.55cm} \caption{Growth rate of the \textbf{left-hand
resonant ion-cyclotron instability}, in the case of an
alpha-particle beam plasma. \textbf{Left}: as a function of the
angle of propagation $\theta$ and the normalized wave number
$kV_{Ap}/\Omega_{p}$ and for an alpha beam density
$n_{0\alpha}=0.1n_{0p}$ and velocity
$v_{0\alpha}=320~\textrm{km/s}=1.8\textrm{V}_{Ap}$.
\textbf{Middle}: as a function of $\theta$ and $n_{0\alpha}$ and for
$v_{0\alpha}=320~\textrm{km/s}\approx1.8\textrm{V}_{Ap}$ and
$kV_{Ap}/\Omega_{p}=0.5$.
\textbf{Right}: as a function of $\theta$ and $v_{0\alpha}/v_{Ap}$
and for $n_{0\alpha}=0.1n_{0p}$ and $kV_{Ap}/\Omega_{p}=0.5$. Here
$T_{0e}=T_{0p}=T_{0\alpha}$.} \label{3D-L-al}
\end{figure*}

For $\theta=15^{\circ}$ (top panel), the instability results from
the intersection between the ion-cyclotron mode ($\omega_{L}$) and
the cyclotron beam mode 2. This region of instability is indicated
by (L) in reference to the left-hand polarization characterizing the
ion-cyclotron mode. Indeed, these two distinct and initially stable
modes merge, within a certain range (corresponding to the red curve)
of the wave number $k$, into one single unstable mode which
satisfies the resonance condition:
\begin{equation}
\omega\approx kv_{0\alpha}~cos(\varphi-\theta)-\Omega_{\alpha}
\label{ano-dop}.
\end{equation}
This condition corresponds to a left-hand resonant cyclotron
excitation of the ion-cyclotron mode through the anomalous Doppler
effect \citep[see,][]{Mecheri:Gary}. As shown in the left panel of
Fig.~\ref{3D-L-al}, this instability extends from smaller $k$, at
propagation angles around $\theta\approx60^{\circ}$ with small
growth rate, i.e. $\gamma\approx0.03\Omega_{p}$, to higher $k$ and
to smaller angles of propagation with a larger growth rate, i.e.
$\gamma\approx0.12\Omega_{p}$. Since the location at x=7.5~Mm and
z=2.2~Mm is characterized by a \textbf{B}$_{0}$-inclination angle
$\varphi=82^{\circ}$, we can therefore say that as $k$ increases
this instability tends to appear at increasingly oblique propagation
angles with respect to the ambient magnetic field, and its growth
rate tends to maximize for perpendicular propagation.
On the middle panel of Fig.~\ref{3D-L-al} the left-hand resonant
instability is shown as a function of $\theta$ and $n_{0\alpha}$,
for the case of $v_{0\alpha}=320$~km/s$~\approx1.8V_{Ap}$ and a
normalized wave number $kV_{Ap}/\Omega_{p}=0.5$. It is clearly seen
that the growth rate of the instability increases with increasing
$n_{0\alpha}$ and gradually is also covering a wider range of
$\theta$, but it stays below approximately
$\theta\approx60^{\circ}$. The maximum growth rate is
$\gamma\approx0.1\Omega_{p}$ for $n_{0\alpha}=0.2n_{0p}$ and
$\theta\approx25^{\circ}$.
This instability is also presented on the right panel of
Fig.~\ref{3D-L-al} as a function of $\theta$ and $v_{0\alpha}$ and
for $n_{0\alpha}/n_{0p}=0.1$ and $kV_{Ap}/\Omega_{p}=0.5$. It is
clearly seen that the instability has a threshold in the beam
velocity $v_{0\alpha}$ below which it does not occur. This threshold
depends on $\theta$ and increases from $v_{0\alpha}\approx 1.5
\textrm{V}_{Ap}$ at $\theta \approx 60^{\circ}$ to
$v_{0\alpha}\approx 2 \textrm{V}_{Ap}$ for quasi-perpendicular
propagation $\theta\approx0^{\circ}$ (knowing that the inclination
angle of \textbf{B}$_{0}$ is $\varphi\approx 82^{\circ}$).

On the other hand, for $v_{0\alpha} \neq 0$ and an angle of
propagation $\theta=65^{\circ}$ (bottom of Fig.~\ref{2D-LR-al}), the
results show the disappearance of the left-hand resonant instability
involving the ion-cyclotron mode and the appearance of another kind
of instability involving the right-handed polarized fast mode, from
which the name right-hand resonant instability is derived. This
instability is indicated by (R) and results from the intersection of
the fast mode ($\omega_{R}$) with the cyclotron beam mode~2
satisfying the resonance condition (\ref{ano-dop}).
\begin{figure*}
\begin{center}
\includegraphics[width=16.cm]{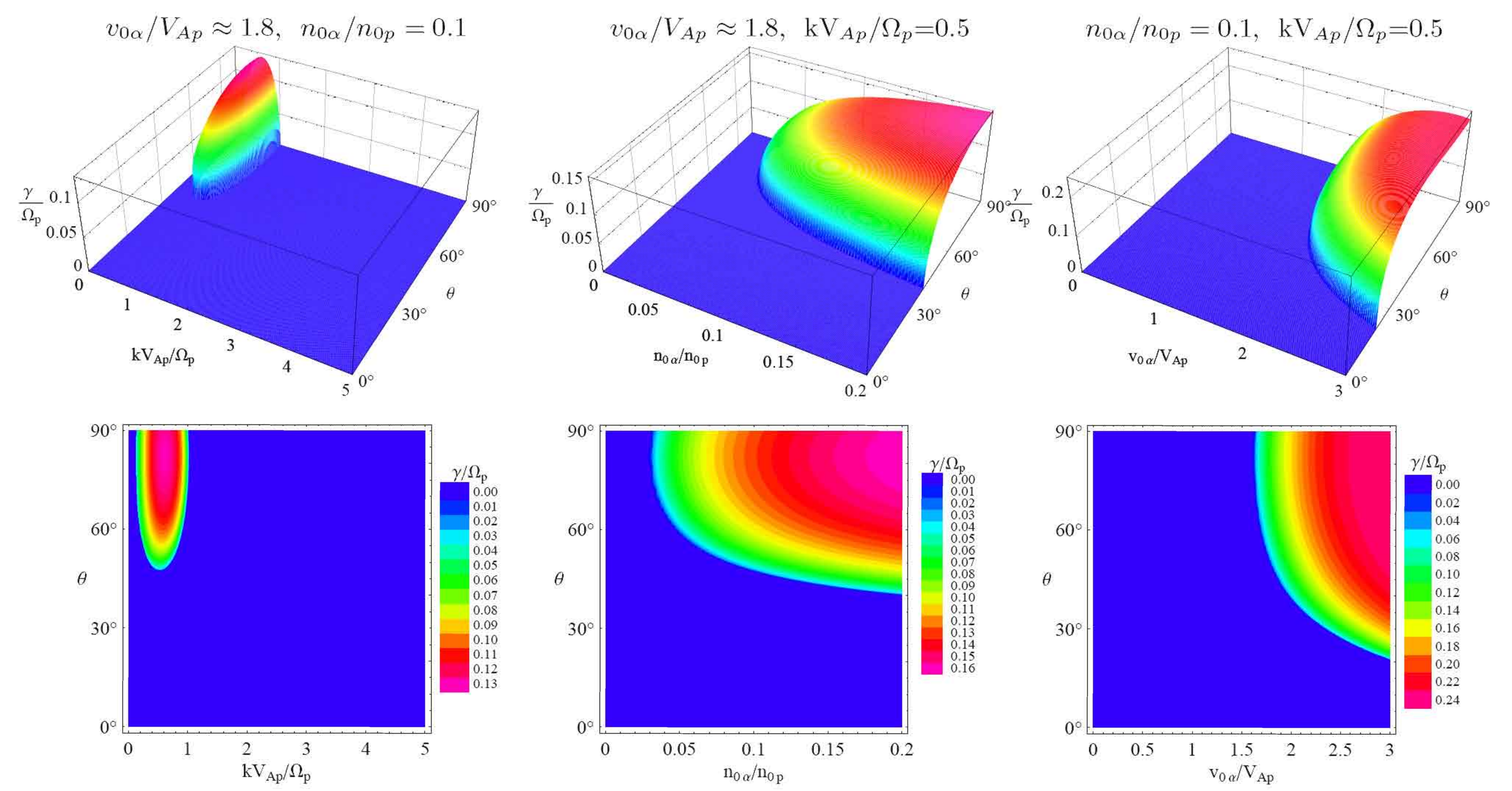}
\end{center}
\vspace{-0.55cm} \caption{Growth rate of the \textbf{right-hand
resonant ion-cyclotron instability}, in the case of an
alpha-particle beam plasma. \textbf{Left}: as a function of the
angle of propagation $\theta$ and the normalized wave number
$kV_{Ap}/\Omega_{p}$ and for an alpha beam density
$n_{0\alpha}=0.1n_{0p}$ and velocity
$v_{0\alpha}=320~\textrm{km/s}=1.8\textrm{V}_{Ap}$.
\textbf{Middle}: as a function of $\theta$ and $n_{0\alpha}$ and for
$v_{0\alpha}=320~\textrm{km/s}\approx1.8\textrm{V}_{Ap}$ and
$kV_{Ap}/\Omega_{p}=0.5$.
\textbf{Right}: as a function of $\theta$ and $v_{0b}/v_{Ap}$ and
for $n_{0\alpha}=0.1n_{0p}$ and $kV_{Ap}/\Omega_{p}=0.5$. Here
$T_{0e}=T_{0p}=T_{0\alpha}$.} \label{3D-R-al}
\end{figure*}
As shown on the left panel of Fig.~\ref{3D-R-al} this instability is
mainly centered around a normalized wave number $kV_{Ap}/\Omega_{p}
\approx 0.5$, and has maximum growth rate
$\gamma\approx0.13\Omega_{p}$ at a large angle of propagation
$\theta\approx90^{\circ}$. The right-hand resonant instability
vanishes for highly oblique propagation with respect to the ambient
field (knowing that the inclination of \textbf{B}$_{0}$ in this
location is $\varphi\approx 82^{\circ}$), and the wave amplitude
grows strongly for decreasing obliquity of the propagation angle.
On the middle panel of Fig.~\ref{3D-R-al}, the right-hand resonant
instability is shown as a function of $\theta$ and $n_{0\alpha}$,
for the case of $v_{0\alpha}=320~\textrm{km/s}\approx 1.8V_{Ap}$ and
$kV_{Ap}/\Omega_{p}=0.5$. It is clearly seen that the growth rate of
this instability increases with increasing $n_{0\alpha}$. This
behavior is more pronounced for higher $\theta$, which corresponds
to decreasingly oblique propagation. The maximum growth rate,
$\gamma\approx0.16\Omega_{p}$, is obtained for
n$_{0\alpha}$=0.2n$_{0p}$ and $\theta\approx80^{\circ}$. We can also
notice that the instability fades away for small propagation angles,
with $\theta\lesssim40^{\circ}$, which corresponds to
quasi-perpendicular propagation (with respect to the field).
In the same figure (on the right panel) we also present the
dependence of this instability upon $\theta$ and $v_{0\alpha}$ for
$n_{0\alpha}/n_{0p}=0.1$ and $kV_{Ap}/\Omega_{p}=0.5$. It can be
seen, similarly to the left-hand instability, that this instability
has a threshold in the beam velocity $v_{0\alpha}$, below which it
does not occur. This threshold depends on $\theta$ and increases
from $v_{0\alpha}\approx 1.6 \textrm{V}_{Ap}$ at large angle of
propagation, $\theta \approx 90^{\circ}$ (quasi-parallel propagation
since $\varphi=82^{\circ}$), to $v_{0\alpha}\approx 2.5
\textrm{V}_{Ap}$ at $\theta\approx25^{\circ}$ (quasi-perpendicular
propagation).

\subsection{Non-local stability analysis}

In this section we intend to go beyond the local treatment of the
waves and perform a non-local wave study using the ray-tracing
equations. The ray-tracing equations are solved employing the
initial conditions obtained from the local solutions of the
dispersion relation (\ref{dispersion}) at the location with
$x_{0}=7.5$~Mm and $z_{0}=2.2$~Mm. We consider the case of an
alpha-particle (He$^{2+}$) beam plasma configuration with a constant
concentration, $n_{\alpha}=0.1n_{p}$, and a constant beam velocity
of $v_{0\alpha}=320~\textrm{km/s}$.

\begin{figure*}
\begin{center}
$\begin{array} {c@{\hspace{0.05in}}c}
\multicolumn{1}{c}{~~~~~~~~~~\mbox{\Large
$x_{0}=7.5$~Mm,~$z_{0}=2.2$~Mm,~$\varphi_{0}\approx82^{\circ}$}}
\\[0.1cm]
\includegraphics[width=17cm]{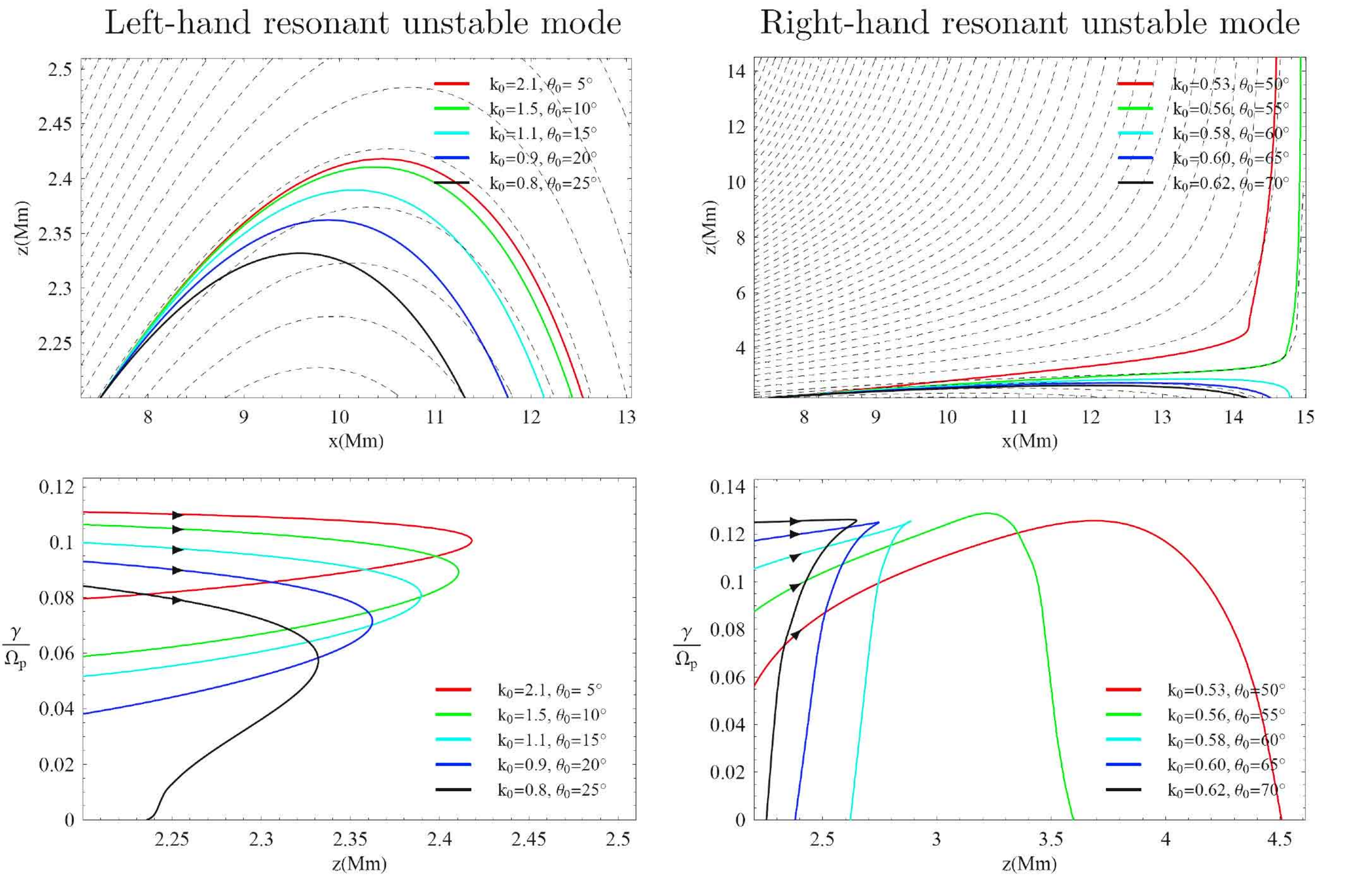}
\end{array}$
\vspace{-0.55cm}
\end{center}
\caption{The ray trajectory (top) and growth rate (bottom) of the
left-hand (left) and the right-hand (right) resonant ion-cyclotron
instabilities in the coronal funnel. These unstable modes are
launched at the initial location ($x_{0}=7.5$~Mm and $z_{0}=2.2$~Mm)
in the funnel (with a \textbf{B}$_{0}$-inclination angle
$\varphi_{0}\approx82^{\circ}$), for different initial angle of
propagation $\theta_{0}$ and wave number $k_{0}$ chosen at the
maximum instability growth rate. The dashed lines represent the
magnetic field lines in the funnel.} \label{RTb-xz-Im}
\end{figure*}

The ray paths of the unstable waves as well as the variation of
their growth rates as a function of height $z$, when the wave is
launched at the initial location ($x_{0}=7.5$~Mm and $z_{0}=2.2$~Mm)
that is characterized by a strong inclination angle
($\varphi\approx82^{\circ}$) of the magnetic field with respect to
the normal on the solar surface, are illustrated in
Fig.~\ref{RTb-xz-Im}. The results are presented for a different
initial angle of propagation, $\theta_{0}$, with which an initial
wave number $k_{0}$ (that is normalized to $\Omega_{p}/V_{Ap}$) is
associated and chosen as to correspond to the maximum growth rate,
$\gamma_{max}$.

Our results show that the ray path of the left-hand unstable wave
(Fig.~\ref{RTb-xz-Im}, on the left) is strongly affected by the
closed-field geometry characterizing this funnel region. Indeed,
this unstable wave starting from its initial position propagates
upward in the coronal funnel to a certain height, where it turns
down again and starts propagating downward to return back to the
initial height. The associated instability growth rate decreases
along that ray path. Since the direction of the group velocity is
always parallel to the ray path and indicates where the energy is
transported, we can say that the energy associated with the
left-hand resonant instability does not reach high altitudes in the
funnel. The smaller $\theta_{0}$ is the higher up this unstable wave
propagates in the funnel.

The right-hand unstable wave (Fig.~\ref{RTb-xz-Im}, on the right) is
also found to be well guided and, depending on $\theta_{0}$, can
follow both closed and open coronal field lines, which may exist
side by side in this region of the funnel. Indeed, for
$\theta_{0}=50^{\circ}$ and $\theta_{0}=55^{\circ}$ the unstable
wave propagates along the open magnetic field lines and reaches high
altitudes in the funnel up to 15 Mm, while for
$\theta_{0}=60^{\circ}$, $65^{\circ}$ and $70^{\circ}$, similarly to
the left-hand instability, the right-hand resonant unstable waves
are affected by the closed-field geometry and reflected back towards
lower altitudes in the funnel. Thus, the energy associated with this
instability is also transported along the magnetic field lines, but
eventually to much greater altitude, i.e. z = 15 Mm, in the funnel
as compared to the left-hand instabilities. For
$\theta_{0}=50^{\circ}$ and $\theta_{0}=55^{\circ}$ the growth rate
of the right-hand instability first increases until respectively
altitudes of z$~\approx~$3.25 Mm and z$~\approx~$3.75 Mm where it
starts to sharply decrease and cancels respectively at
z$~\approx~$3.6 Mm and z$~\approx~$4.5 Mm. On the other hand, for
$\theta_{0}=60^{\circ}$, $65^{\circ}$ and $70^{\circ}$, the growth
is first slightly increases during the upward propagation phase, and
then it rapidly decreases to a zero value while the wave is
propagating downward.

\section{Conclusion}\label{conclusion}

We have studied beam-driven electromagnetic instabilities near the
ion-cyclotron frequency in a coronal funnel using the multi-fluid
model. We have considered the case of an alpha-particle beam
propagating in the funnel parallel to the ambient magnetic field
lines. In agreement with kinetic dispersion theory, the local
solutions of the dispersion relation revealed the presence of two
kinds of instabilities: the left-hand and right-hand resonant
instabilities. The left-hand and right-hand instabilities arise from
the resonant excitation of, respectively, the left-hand-polarized
ion-cyclotron mode and right-hand-polarized fast mode through the
anomalous Doppler effect, see Eq. (\ref{ano-dop}). For the studied
coronal region, our results indicate that the left-hand resonant
instability develops for strongly oblique wave propagation with
respect to the ambient magnetic field, with a maximum growth rate at
a quasi-perpendicular propagation, and disappears for weakly oblique
propagation. Oppositely, the right-hand instability develops for a
weakly-oblique propagation to the ambient magnetic field, with a
maximum growth rate at quasi-parallel propagation, and ceases for
highly oblique or quasi-perpendicular propagation.

The nonlocal ray-tracing analysis revealed that both instabilities
are sensitively affected by the magnetic field geometry and found to
propagate closely along the field lines. The left-hand resonant
instability is rapidly reflected, thus obeying the constraints
imposed by a closed-field configuration. The associated growth rate
slightly decreases and eventually cancels along the ray path. On the
other hand, the right-handed resonant instabilities are also very
well guided along the magnetic field lines of the funnel. This
instability, for small initial angles of propagation, appears to
follow the open field lines and can propagate higher up in the
funnel, yet with a rapidly decreasing growth rate.

Consequently, fast ion beams in the magnetically open corona can
provide enough energy for driving micro-instabilities through
resonant wave-particle interactions. These instabilities may
constitute in turn an important energy source for high-frequency
ion-cyclotron waves which have been invoked to play a relevant role
in the heating of coronal ions through cyclotron damping.
%
%
\bibliographystyle{aa}
\bibliography{Mecheri-beam}

\begin{thebibliography}{28}
\expandafter\ifx\csname natexlab\endcsname\relax\def\natexlab#1{#1}\fi

\bibitem[{{Axford} \& {McKenzie}(1992)}]{Mecheri:Axford92}
{Axford}, W.~I. \& {McKenzie}, J.~F. 1992, in Solar Wind Seven Colloquium, ed.
  E.~{Marsch} \& R.~{Schwenn}, 1--5

\bibitem[{{Bernstein} \& {Friedland}(1984)}]{Mecheri:Bernstein}
{Bernstein}, I.~B. \& {Friedland}, L. 1984, in Basic Plasma Physics: Handbook
  of Plasma Physics, Volume 1, ed. A.~A. {Galeev} \& R.~N. {Sudan}, 367

\bibitem[{{Brekke} {et~al.}(1997){Brekke}, {Kjeldseth-Moe}, {Brynildsen},
  {Maltby}, {Haugan}, {Harrison}, {Thompson}, \& {Pike}}]{Mecheri:Brekke}
{Brekke}, P., {Kjeldseth-Moe}, O., {Brynildsen}, N., {et~al.} 1997, \solphys,
  170, 163

\bibitem[{{Cap}(1978)}]{Mecheri:Cap}
{Cap}, F.~F. 1978, {Handbook on plasma instabilities. Volume 2} (New York,
  Academic Press, 1978.~572 p.)

\bibitem[{{Cranmer} {et~al.}(1999){Cranmer}, {Field}, \&
  {Kohl}}]{Mecheri:Cranmer}
{Cranmer}, S.~R., {Field}, G.~B., \& {Kohl}, J.~L. 1999, \apj, 518, 937

\bibitem[{{Feldman} {et~al.}(1996){Feldman}, {Barraclough}, {Phillips}, \&
  {Wang}}]{Mecheri:Feldman}
{Feldman}, W.~C., {Barraclough}, B.~L., {Phillips}, J.~L., \& {Wang}, Y.-M.
  1996, \aap, 316, 355

\bibitem[{{Fontenla} {et~al.}(1993){Fontenla}, {Avrett}, \&
  {Loeser}}]{Mecheri:Fontenla}
{Fontenla}, J.~M., {Avrett}, E.~H., \& {Loeser}, R. 1993, \apj, 406, 319

\bibitem[{{Gabriel}(1976)}]{Mecheri:Gabriel}
{Gabriel}, A.~H. 1976, Royal Society of London Philosophical Transactions
  Series A, 281, 339

\bibitem[{{Gary}(1993)}]{Mecheri:Gary}
{Gary}, S.~P. 1993, {Theory of Space Plasma Microinstabilities} (UK: Cambridge
  University Press.)

\bibitem[{{Hackenberg} {et~al.}(2000){Hackenberg}, {Marsch}, \&
  {Mann}}]{Mecheri:HackenbergB}
{Hackenberg}, P., {Marsch}, E., \& {Mann}, G. 2000, \aap, 360, 1139

\bibitem[{{Hollweg}(2000)}]{Mecheri:Hollweg00}
{Hollweg}, J.~V. 2000, \jgr, 105, 15699

\bibitem[{{Hollweg} \& {Isenberg}(2002)}]{Mecheri:Hollweg02}
{Hollweg}, J.~V. \& {Isenberg}, P.~A. 2002, Journal of Geophysical Research
  (Space Physics), 107, 12

\bibitem[{{Innes} {et~al.}(1997){Innes}, {Inhester}, {Axford}, \&
  {Willhelm}}]{Mecheri:Innes}
{Innes}, D.~E., {Inhester}, B., {Axford}, W.~I., \& {Willhelm}, K. 1997, \nat,
  386, 811

\bibitem[{{Isenberg} {et~al.}(2000){Isenberg}, {Lee}, \&
  {Hollweg}}]{Mecheri:Isenberg00}
{Isenberg}, P.~A., {Lee}, M.~A., \& {Hollweg}, J.~V. 2000, \solphys, 193, 247

\bibitem[{{Kohl} {et~al.}(1997){Kohl}, {Noci}, {Antonucci}, {Tondello},
  {Huber}, {Gardner}, {Nicolosi}, {Strachan}, {Fineschi}, {Raymond}, {Romoli},
  {Spadaro}, {Panasyuk}, {Siegmund}, {Benna}, {Ciaravella}, {Cranmer},
  {Giordano}, {Karovska}, {Martin}, {Michels}, {Modigliani}, {Naletto},
  {Pernechele}, {Poletto}, \& {Smith}}]{Mecheri:Kohl}
{Kohl}, J.~L., {Noci}, G., {Antonucci}, E., {et~al.} 1997, \solphys, 175, 613

\bibitem[{{Li} {et~al.}(1999){Li}, {Habbal}, {Hollweg}, \&
  {Esser}}]{Mecheri:Li99}
{Li}, X., {Habbal}, S.~R., {Hollweg}, J.~V., \& {Esser}, R. 1999, \jgr, 104,
  2521

\bibitem[{{Markovskii}(2001)}]{Mecheri:Markovskii01}
{Markovskii}, S.~A. 2001, \apj, 557, 337

\bibitem[{{Markovskii} \& {Hollweg}(2004)}]{Mecheri:Markovskii04}
{Markovskii}, S.~A. \& {Hollweg}, J.~V. 2004, Nonlinear Processes in
  Geophysics, 11, 485

\bibitem[{{Marsch} \& {Tu}(2001)}]{Mecheri:Marsch01}
{Marsch}, E. \& {Tu}, C.-Y. 2001, \jgr, 106, 227

\bibitem[{{Mecheri} \& {Marsch}(2006)}]{Mecheri:Mecheri}
{Mecheri}, R. \& {Marsch}, E. 2006, Royal Society of London Philosophical
  Transactions Series A, 364, 537

\bibitem[{{Ofman} {et~al.}(2002){Ofman}, {Gary}, \&
  {Vi{\~n}as}}]{Mecheri:Ofman02}
{Ofman}, L., {Gary}, S.~P., \& {Vi{\~n}as}, A. 2002, Journal of Geophysical
  Research (Space Physics), 107, 9

\bibitem[{{Schrijver} {et~al.}(1998){Schrijver}, {Title}, {Harvey}, {Sheeley},
  {Wang}, {van den Oord}, {Shine}, {Tarbell}, \&
  {Hurlburt}}]{Mecheri:Schrijver98}
{Schrijver}, C.~J., {Title}, A.~M., {Harvey}, K.~L., {et~al.} 1998, \nat, 394,
  152

\bibitem[{{Stix}(1992)}]{Mecheri:Stix}
{Stix}, T.~H. 1992, {Waves in Plasmas} (New York: American Institute of
  Physics, 1992)

\bibitem[{{Tu} {et~al.}(2004){Tu}, {Marsch}, \& {Qin}}]{Mecheri:Tu04}
{Tu}, C.-Y., {Marsch}, E., \& {Qin}, Z.-R. 2004, Journal of Geophysical
  Research (Space Physics), 109, 5101

\bibitem[{{Vocks} \& {Marsch}(2001)}]{Mecheri:Vocks01}
{Vocks}, C. \& {Marsch}, E. 2001, \grl, 28, 1917

\bibitem[{{Voitenko} \& {Goossens}(2002)}]{Mecheri:Voitenko02}
{Voitenko}, Y. \& {Goossens}, M. 2002, \solphys, 206, 285

\bibitem[{{Weinberg}(1962)}]{Mecheri:Weinberg}
{Weinberg}, S. 1962, Physical Review, 126, 1899

\bibitem[{{Xie} {et~al.}(2004){Xie}, {Ofman}, \& {Vi{\~n}as}}]{Mecheri:Xie04}
{Xie}, H., {Ofman}, L., \& {Vi{\~n}as}, A. 2004, Journal of Geophysical
  Research (Space Physics), 109, 8103

\end{thebibliography}
\end{document}